\documentclass[]{spie}  

\usepackage{amsmath,amsfonts,amssymb,siunitx,graphicx, soul}
\usepackage[colorlinks=true, allcolors=blue]{hyperref}

\title{Overview of the SAPHIRA Detector for AO Applications}

\author[a,b]{Sean B. Goebel}
\author[a]{Donald N.B. Hall}
\author[b,c,d,e]{Olivier Guyon}
\author[a]{Eric Warmbier}
\author[a]{Shane M. Jacobson}
\affil[a]{Institute for Astronomy, University of Hawaii, 640 North A'ohoku Place, Hilo, HI 96720}
\affil[b]{Subaru Telescope, National Astronomical Observatory of Japan, 650 North A'ohoku Place, Hilo, HI 96720}
\affil[c]{Steward Observatory, University of Arizona, Tucson, AZ 85721}
\affil[d]{College of Optical Sciences, University of Arizona, Tucson, AZ 85721}
\affil[e]{Astrobiology Center of NINS, 2-21-1, Osawa, Mitaka, Tokyo, 181-8588, Japan}

\authorinfo{Send correspondence to Sean Goebel, sgoebel@ifa.hawaii.edu}

\begin{document}
\maketitle

\begin{abstract}
We discuss some of the unique details of the operation and behavior of Leonardo SAPHIRA detectors,
particularly in relation to their usage for adaptive optics wavefront sensing. SAPHIRA 
detectors are 320$\times$256@24 $\mu$m pixel HgCdTe linear avalanche photodiode arrays and are sensitive to
0.8-2.5 $\mu m$ light. SAPHIRA arrays
permit global or line-by-line resets, of the entire detector or just subarrays of it, and the order
in which pixels are reset and read enable several readout schemes. We discuss three readout modes,
the benefits, drawbacks, and noise sources of each, and the observational modes for which each is 
optimal. We describe the ability of the detector to read subarrays for increased frame rates, and
finally clarify the differences between the avalanche gain (which is user-adjustable) and the 
charge gain (which is not). 
\end{abstract}



\section{INTRODUCTION}
Due to their ability to detect individual photons with both high temporal and spatial resolutions, 
electron-multiplying CCDs (EMCCDs) have greatly improved the sensitivity limits of adaptive optics (AO) 
systems. 
However, EMCCDs are sensitive to optical wavelengths, and in most AO implementations, the
science instrument operates at near-infrared wavelengths. This wavelength sensitivity difference 
between the wavefront sensor and science instrument occurs for two reasons. First, the 
path length differences introduced by atmospheric turbulence are approximately independent of
wavelength, so the aberrations as a fraction of phase are reduced at longer wavelengths. 
Therefore, better image quality is obtained at longer wavelengths, and so the science
instruments are designed to take advantage of this. Second,
until now there have not existed high frame rate, low noise,
reasonable cost infrared detector arrays, so it was not feasible to do high-order wavefront 
sensing at
near-infrared wavelengths. The Selex Avalanche Photodiode for HgCdTe InfraRed Array (SAPHIRA)
detector is the first such technology that enables this.

Operating the wavefront sensor at near-infrared wavelengths provides several 
benefits~\cite{Wizinowich2016}. It has the potential to expand the sky coverage available for 
natural guide
star sensing because it enables the observation of targets that have minimal emission at optical 
wavelengths, such as late-type stars or obscured objects. In particular, there is significant 
interest in observing nearby
M dwarfs because they have favorable contrasts for directly imaging reflected-light extrasolar 
planets located in their habitable zones.
Second, if both the science module and wavefront sensor are sensitive to similar
wavelengths, they can share a greater fraction of optical elements inside the instrument and thereby 
minimize noncommon path errors. For these reasons, SAPHIRA detectors enable exciting new AO 
observing modes.

SAPHIRA arrays are well-suited for wavefront sensing because they provide two advantages over other
infrared devices. First, they have a user-adjustable avalanche gain. This is
discussed at length in Section~\ref{sec:gains}, but in short, it amplifies the signal from photons
but not read noise sources. At high gains, the signal of an individual photon is greater than the
noise. If the noise is divided by the avalanche gain, it can be 
$<$1 $e^-$. The second advantage of SAPHIRA is that all outputs are still used when reading 
subarrays, so greatly increased frame rates can be achieved by reading less than the full detector
(Section~\ref{sec:subwindows}). This is in contrast to (for example) Teledyne HAWAII detectors, in 
which each output reads a contiguous stripe of the detector and only a single output is used
when in sub-array mode.

SAPHIRA detectors are 320$\times$256 pixel mercury cadmium telluride arrays and have a 24 $\mu m$
pixel pitch~\cite{Baker2016}. They are manufactured by Leonardo (formerly called Selex)
and have 32 outputs. The absorption layer (where photons are intended to be absorbed) has 0.8-2.5 
$\mu m$ sensitivity
and is transparent at longer wavelengths. These longer wavelengths must be filtered out to avoid 
spurious signal from
photons absorbed in the multiplication layer. The multiplication layer, which is where the avalanche
multiplications occur, lies below the absorption layer and has sensitivity to 3.5 
$\mu m$~\cite{Finger2016}.
Because SAPHIRAs are a developmental program, they are assigned 
mark numbers which refer to the generation of metalorganic vapour phase epitaxy (MOVPE)
layer architecture; higher numbers are newer designs. Mk. 3 SAPHIRA arrays were the first
science-grade ones and were delivered in 2013. The Mk. 13 and 14 arrays (which are indistinguishable 
for the purposes of wavefront sensing) were delivered in 2015 and (as of the time of writing) remain 
the best arrays for telescope deployments due to their uniform cosmetic qualities. Later generations
of arrays have improved in some areas (e.g. dark current) but had problems in other areas (e.g.
dead pixels or operational reliability). There are plans to produce larger format SAPHIRA arrays
for low-background imaging and high-order wavefront sensing on thirty-meter-class 
telescopes~\cite{Hall2016A}.
SAPHIRA devices produced until late 2015 utilized ME911 read out integrated circuits (ROICs), 
which only permitted global resets and therefore were limited to up-the-ramp readout mode 
(Section~\ref{sec:utr}). The ME1000 ROIC enabled row-by-row resets, which made 
possible read-reset-read and read-reset modes (Sections~\ref{sec:rrr} and~\ref{sec:rr}).
The ME1001 ROIC is functionally identical to the ME1000 ROIC, but it features reduced
glow during operation.

While the operation of the SAPHIRA APD array is very similar to that of conventional HgCdTe arrays such
as the Teledyne HAWAII series~\cite{Beletic2008} and the Raytheon Virgo~\cite{Starr2016}, there are some
important differences. The conventional arrays are normally operated at a bias of a few hundred $mV$, 
occasionally up to a volt, and the full bias voltage between the substrate and the node is within the 
dynamic range of the readout circuit. After reset, the array discharges to the substrate voltage. The 
depletion is modest and varies with
bias voltage, and that results in a highly bias-dependent diode capacitance.
In contrast, the SAPHIRA typically operated at bias voltages of $2.5-18\;V$, way beyond the dynamic range of 
the readout circuit, and the node voltage is referenced to ground in the ROIC. At these bias voltages the 
photodiode is largely or completely depleted, and the response is highly linear over the dynamic range. As 
a result of the depletion, there is no trapping of charge when the bias voltage is changed.

SAPHIRA arrays are beginning to enter widespread deployments. The University of Hawaii Institute for
Astronomy has extensively tested the detectors in the lab and at telescopes. Over the last several
years, we have continuously deployed a SAPHIRA system to the SCExAO extreme AO instrument at Subaru 
Telescope, where it is primarily used for focal-plane wavefront 
sensing~\cite{Goebel2016}. It has also been deployed~\cite{Atkinson2014} short-term to the NASA 
Infrared Telescope Facility (IRTF), where it obtained diffraction-limited images of binary star systems using
the lucky imaging\cite{Fried1978} technique. Also, a SAPHIRA system is currently being integrated into
the Keck II AO system in order to enable near-infrared 
pyramid wavefront sensing~\cite{Mawet2016}. SAPHIRA is being used for tip/tilt
sensing on the Robo-AO instrument, which was initially deployed to the 1.5 $m$ telescope at 
Palomar~\cite{Baranec2015} and then 
later to the 2.1 $m$ telescope at Kitt Peak~\cite{Jensen-Clem2018}. While at Robo-AO, SAPHIRA
produced its first science publication~\cite{Han2017}. ESO has
also been evaluating SAPHIRA arrays~\cite{Finger2012}, and they are deployed to the Very
Large Telescope's GRAVITY 
instrument for wavefront sensing at the individual telescopes and fringe tracking at the combined 
focus~\cite{Mehrgan2016}. Lastly, First Light Imaging has developed a commercial system called
C-RED One that combines the SAPHIRA detector, camera and associated cooling, and readout
electronics~\cite{Greffe2016}.

In this paper, we report on the operation modes and intricacies of behavior of SAPHIRAs. 
Unless stated otherwise, the data presented below were collected at a temperature of 85 K
and at unity avalanche gain (i.e. a bias voltage of 2.5 $V$) with Mark 13 and 14 arrays on
ME1000 ROICs.

\section{Readout Modes}\label{sec:readoutmodes}
The readout modes available on SAPHIRA detectors differ according to when and the number of times that
pixels are read between resets. 
We have operated the SAPHIRA in three different readout
schemes: 1) sampling up-the-ramp mode, which utilizes an initial reset of all pixels
followed by multiple reads; 2) read-reset-read mode, which reads a row of pixels, resets it, and reads
it again before clocking to the next row; and 3) read-reset mode, whereby a row is read and
then reset before moving to the next row, and a reference image is subtracted in order to remove 
the pedestal voltage (sometimes called the fixed-pattern noise or bias). These readout modes are 
independent of subarray size. In practice, we use read-reset mode for AO-related observations because
it provides regular time sampling (unlike sample up-the-ramp mode) and better noise performance
and double the frame rate compared to read-reset-read mode.
In the following sections, we discuss each readout mode in turn. 

\subsection{Sampling Up-the-ramp Mode}\label{sec:utr}
In sampling up-the-ramp (SUTR) mode, the entire array is reset, and then it is non-destructively read a 
user-specified number of times~\cite{Chapman1990}. These reads can be processed in one of two ways. 
First, the user can fit a line to the ``ramp'' (flux vs. time) behavior for each pixel. This removes
the pedestal voltage (fixed-pattern noise) and reduces the white read noise variance as $1/N$, where 
$N$ is the number of times the array is read per ramp. 
A mathematical analysis of SUTR (and the conceptually similar Fowler sampling)
is given by Garnett and Forrest (1993)~\cite{Garnett1993}. Additionally, if a pixel saturates partway 
through the ramp due to high flux or a cosmic ray 
strike, the reads in which the pixel is saturated can be excluded from the fit. 
When SUTR data are reduced by line-fitting, one ``science'' frame is produced for each ramp. This is
well-suited for low-background observations in which read noise dominates and the exposures are long, but
it is poorly suited for instances where fast analysis of the images is required (such as AO wavefront
sensing).

The second way to reduce SUTR data is to subtract subsequent reads. If being used for AO wavefront
sensing, this enables more frequent updates to the correction. By subtracting temporally adjacent frames,
one removes the pedestal voltage and $kTC$ (reset level) noise because it is present in both images,
leaving only the flux that accumulated in the pixel between those two exposures and the read noise.
However, this mode causes three challenges. First, it requires that the dynamic range
be rationed across the ramp (if there are $N$ reads per ramp, then no individual read can utilize more
than $1/N$ of the dynamic range, or else the final reads of the pixel will contain no new flux). 
Second, there is irregular time sampling because it wouldn't make sense 
to subtract the read immediately before a reset from the one immediately after the reset because
this would result in negative flux and $kTC$ noise.
This is problematic for applications that require regular temporal sampling.
Third, due to the intrinsic nonlinearity of infrared
arrays, the responsivity of the detector decreases as the pixels accumulate flux. This causes frame pairs 
toward the end of the ramp to incorrectly report less flux than an equally-illuminated pair at the 
beginning of the ramp. One can compensate for this last effect with a standard linearity correction, 
however.

Often, the frame read immediately after a reset contains excess noise due to settling effects. In 
practice, we typically exclude the first frame following each reset from analysis. This
exacerbates the irregularity of time sampling in SUTR mode and reduces its duty cycle, particularly 
when there  are few reads between resets. 

In both methods of reducing SUTR data, variations in the voltage of a pixel after one
reset compared to its voltage after another reset has no effect because it is in all reads of 
a given ramp. This is called $kTC$ noise because its variance is given by
\begin{equation}
\sigma_{kTC}^2 = k_{B} T C
\label{eqn:ktc}
\end{equation}
where $k_{B}$ is the Boltzmann constant, $T$ is temperature, and and $C$ is node capacitance. We 
typically operate the detector at $T=85$K, and 
$C=28$fF 
for SAPHIRA, so $\sigma_{kTC}\approx36e^-$. 
Unlike read noise, $kTC$ noise is not affected by the pixel read rate. 
Because SUTR readouts are not affected by $kTC$ noise, they enable the lowest noise of the 
readout modes discussed here. SUTR readouts are best-suited for low-background, long integrations.

\subsection{Read-reset-read Mode}\label{sec:rrr}
In read-reset-read mode, a row is read, reset, and then read again before clocking to the next
one. The read immediately following the reset (which contains only the pedestal voltage) is then 
subtracted from the read preceding the next reset after scanning through the array
(which contains the pedestal voltage plus flux).
Therefore, there are two reads per science frame. This mode removes the effects of $kTC$
noise.

\begin{figure}[!ht]
\begin{centering}
\includegraphics[width=15 cm]{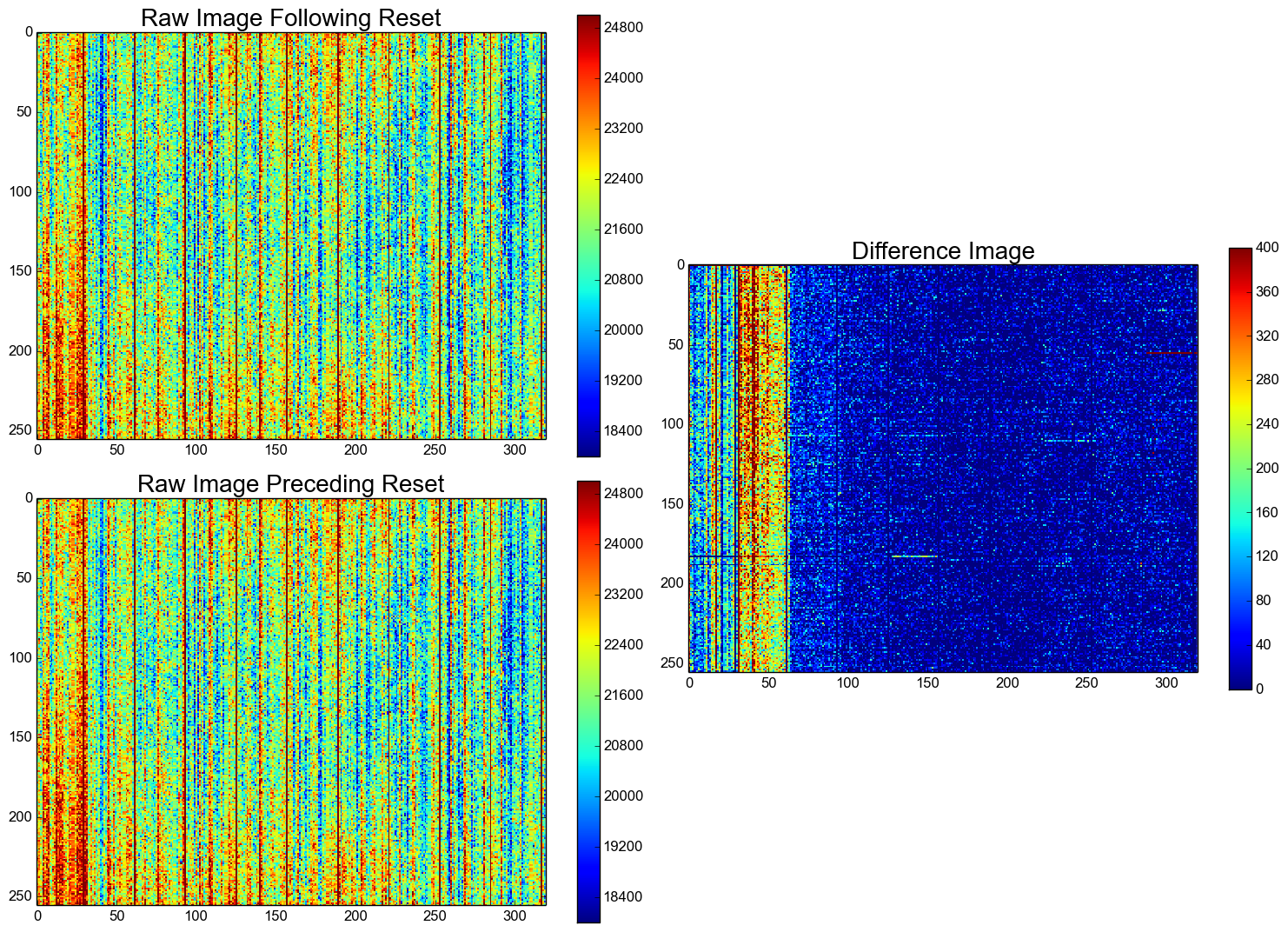}
\caption{On the left are consecutive unilluminated raw frames from SAPHIRA after (top) and before 
(bottom) the reset in read-reset-read mode. Approximately 2.7 ms has elapsed between the two reads. As 
is normal for images from infrared arrays, the pedestal noise dominates both images. However, when the
before image is subtracted from the after image, not all the noise  disappears. An entire
row is reset at a time, the detector clocks from left to right, and the column blocks read immediately
following the reset have spurious flux. Operating the SAPHIRA with a subarray that excludes these left
columns does not solve the problem; the column blocks with spurious flux simply move to the edge of the
subarray being read, since they are read immediately following the reset. All images have units of 
ADUs. We have noticed that this effect is reduced on other SAPHIRA arrays. Because of this post-reset 
setting problem, we do not typically operate SAPHIRA arrays in read-reset-read mode.}
\label{fig:rrr}
\end{centering}
\end{figure}

In practice, read-reset-read mode presents two problems. First, and more importantly, settling effects
immediately following the reset lead to spurious flux in the first column blocks read. This problem is
illustrated in Figure~\ref{fig:rrr}. The amplitude of this effect varies from one SAPHIRA array to 
another, and we theorize that it can be mitigated by inserting a delay between the reset and first read,
though we have not tested this because it would slow the frame rate.
Second, because there is read noise associated with both reads, these add in quadrature for the 
science frame. For the $\sim$100 kHz clocking rates used for most IR detectors, the read noise is much
less than the $kTC$ noise. However, the read noise approximately scales with the square root of the 
sampling frequency, and for the 1-10 MHz sampling frequencies used for SAPHIRA wavefront sensing, 
these two
noise sources become comparable. Because of the spurious flux and replacement of $kTC$ noise with
similar-magnitude read noise, we do not commonly use read-reset-read mode.

\subsection{Read-reset Mode}\label{sec:rr}
If the read noise of one frame is comparable to its $kTC$ noise, then the correlated double sampling
of read-reset-read mode (which produces an image free of $kTC$ noise but containing read noise contributed
by both frames) is no better than subtracting a reference dark frame that has been 
produced from averaging together many frames (and therefore contains negligible $kTC$ and read noise)
from a single read (which does contain these noise sources).
The read noise depends on the sampling rate and specifics of the readout electronics, but 
during typical usage of SAPHIRA, it is of the same order as the 36 $e^-$ $kTC$ noise calculated
from Equation~\ref{eqn:ktc}. 

In read-reset mode, the pedestal voltage of the detector is removed by averaging 
together many unilluminated frames and then subtracting this reference frame from each individual frame.
During science operation, these dark frames can be collected before or after the data frames. However,
a new reference frame needs to be collected for each bias voltage (the bias voltage controls the avalanche 
gain), since the pedestal voltage pattern on the detector changes with bias voltage. Read-reset mode is 
our standard mode for AO observations because its noise
is similar or lower than that of read-reset-read mode, it produces a ``science'' frame for every read,
and the temporal sampling is regular. Additionally, 
read-reset mode does not exhibit the post-reset settling problem (the gradient of spurious flux) of 
read-reset-read mode.

\subsection{Comparison of the Readout Modes}\label{sec:comparison}
A comparison of the loop update rate and noise sources for each of the readout modes is
provided in Table~\ref{table:readoutmodes}.
\begin{table}[h!]
\centering
\begin{tabular}{ |c|c|c|c| } 
 \hline
	& \textbf{Maximum loop}  
    & \textbf{Noise variance}
	& \textbf{Comments} \\
    & \textbf{update rate}   
    & $\sigma_{total}^2=\sigma_{photon}^2+$	
    &		  
    \\ \hline 
      
	\textbf{Up the ramp}	
    & Every read, with 
    & 
    & Dynamic range must be split
    \\     
    (CDS pairs processed	
    & interruption 
    & $2\sigma_{RN}^2 $ 
    & between the reads in the ramp.
    \\
    independently)
    & following reset
    &	
    & 
    \\ \hline 
     
  	\textbf{Up the ramp} 
    & 
    & 
    & Cosmic rays and saturation \\
    (flux determined by
    & Every ramp
    & $\sigma_{RN}^2/N_{reads}$ 
    & can be mitigated by excluding \\
    fitting to ramp) 
    &				
    & 
    & those reads from fitting.
    \\ \hline 
     
	\textbf{Read-reset-read} 
    & Every two reads 
    & $2\sigma_{RN}^2 + \sigma_{spurious}^2$ 
    & The first few column blocks \\ 
    
    & 
    & 
    & may contain spurious flux.
    \\ \hline 
    
	\textbf{Read-reset} 
    & Every read 
    & $\sigma_{RN}^2 + \sigma_{settling}^2 + \sigma_{kTC}^2 $ 
    & Requires reference dark frame.
    \\ \hline 
\end{tabular}
\caption{A comparison of the different readout modes available for the SAPHIRA detector.
$\sigma_{RN}^2$ is the variance in read noise, $N_{reads}$ is the number of reads between resets
in a ramp, $\sigma_{kTC}^2$ is the kTC noise, $\sigma_{spurious}^2$ is the noise associated with 
post-reset settling in read-reset-read mode,
$\sigma_{settling}^2$ is noise due to thermal drifts and is negligible if the data and reference
frames are collected near in time to each other or both are collected after at least two hours of
detector operation. $\sigma_{RN}^2$ is assumed to be white. Noise due to radio frequency 
interference (Section~\ref{sec:rfi}) is a component of $\sigma_{RN}^2$.}
\label{table:readoutmodes}
\end{table}

The SAPHIRA detector can be operated with global resets or line-by-line resets. The user can
configure which regions are reset; one can save time by not resetting portions of the array
not being read. We typically operate the detector with global resets
when sampling up-the-ramp and line-by-line resets in read-reset and read-reset-read mode. However,
line-by-line resets can be used in all three modes. ME1000 and ME1001 ROICs support both global 
and line-by-line resets, whereas ME911 ROICs only support global resets.

Due to the fact that an entire row is reset at once, but then 32 pixels are read in each increment along
the row, in read-reset mode, one side of the detector will see less flux. In other words, for pixels on
one side of the array, there is a greater interval between the read and reset than for pixels on the
opposite side of the array. This effect is worse for a short, fat subarray than for a tall, thin
one. For a 320$\times$256 pixel array, the difference in flux from one side to the other is 
$\sim$0.3\%. For a 320$\times$64
subarray, however, this effect is 1.3\%. For most astronomical applications, a standard 
flat field correction effectively compensates for this effect. This effect is not relevant to
sampling up-the-ramp readouts because one is calculating the slope of the flux between reads, and it
does not affect read-reset-read images because the read of a given pixel is delayed relative to the 
reset an equal amount in both frames.

\begin{figure}[!ht]
\begin{centering}
\includegraphics[width=15 cm]{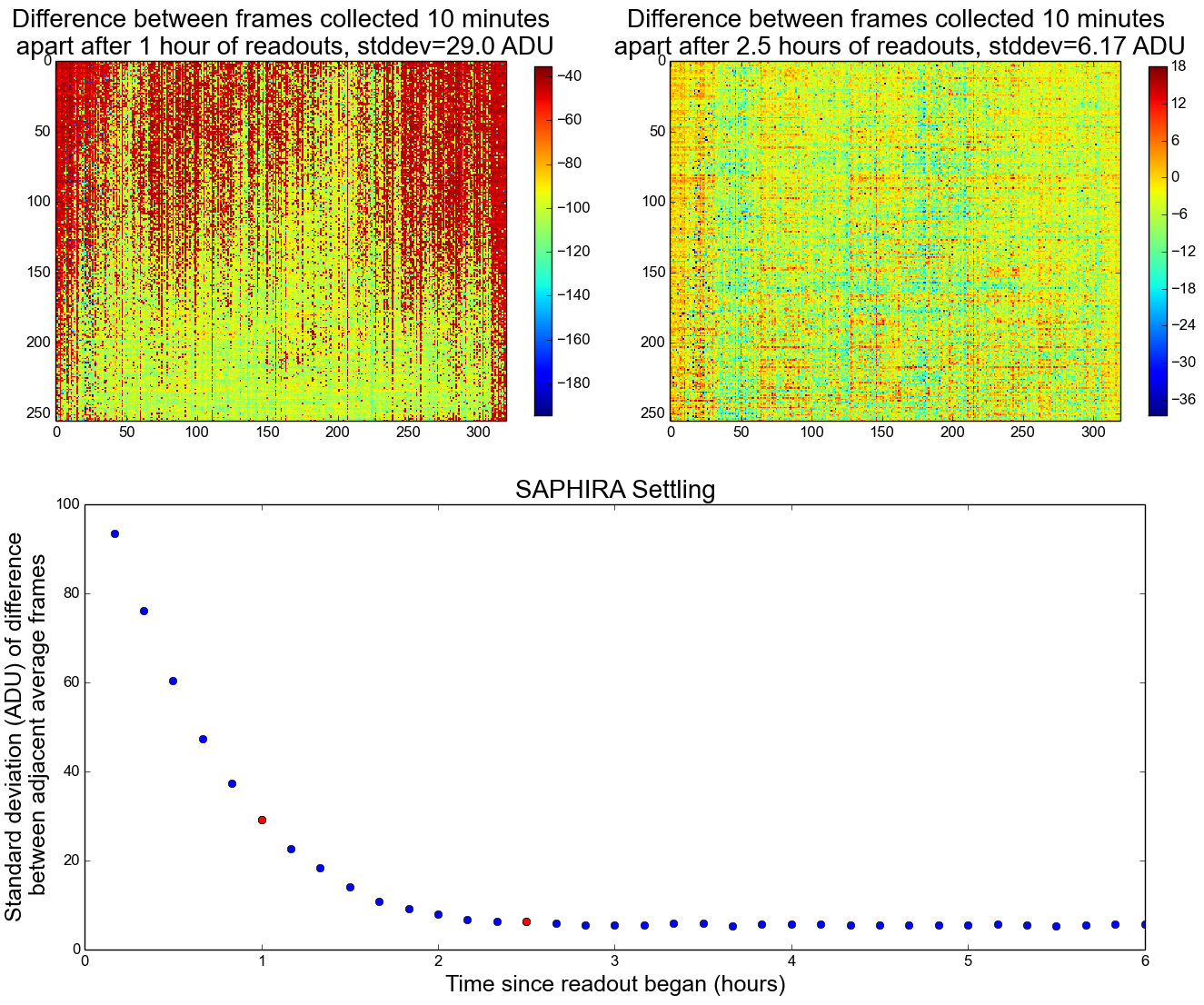}
\caption{While the SAPHIRA continuously streamed images, frames were saved every 10 minutes for
several hours. At each interval, 1000 frames were saved and averaged together in order to reduce
read noise. Each average frame was then subtracted from the average frames collected 10 minutes
earlier and later to see how the pedestal voltage of the detector changed. For
approximately the first two hours of image streaming, the pedestal voltage exhibited
slow drifts (as shown in the top left figure). However, after about 2-2.5 hours, the detector
had largely stabilized (top right). The bottom plot shows how the spatial standard deviation of 
these difference frames decreases with time. The two example frames shown at the top are identified
with red points in the bottom plot. For this reason, it is important to clock the detector for
at least two hours before observations that require good stability.}
\label{fig:settling}
\end{centering}
\end{figure}

In addition to the post-reset settling effects discussed in Sections~\ref{sec:utr} and~\ref{sec:rrr},
we have also observed a smaller amplitude but longer duration thermal settling in the array. When
an array begins clocking after sitting idle, the power dissipation in it changes significantly. This 
causes thermal variations in the array, and it can take hours for these to reach equilibrium. During
this time, pixels exhibit slow drifts. We performed a test in which we read out
an unilluminated array for several hours in read-reset mode. Every ten minutes, we saved 1000 frames 
and averaged them to produce an image of the detector's pedestal voltage with minimal read noise. 
We then subtracted
subsequently collected average frames and computed the standard deviation to see how much the
pedestal voltage drifted. The result is shown in Figure~\ref{fig:settling}. Again, an approximately two-hour 
settling timescale
was observed. For this reason, it is important to run an ``idle sequence'' before observations in which 
no data are saved but the  detector clocks as it would during science operations. The SAPHIRA detector
should be operated like this for at least two hours prior to observations where temporal stability
is required (such as before collecting the reference frames for subsequent read-reset mode 
observations). It is worth noting that clocking-caused thermal drifts are not unique to SAPHIRA arrays;
clocking the same way when idling and observing is key to obtaining good performance from CMOS 
detectors in general.

This same two-hour timescale for SAPHIRA thermal drifts was also observed in dark current 
measurements performed by Atkinson et al.~\cite{Atkinson2017}. While operating SAPHIRA arrays using
up-the-ramp sampling, they noted that it took approximately two hours of continuous clocking before the dark
current stabilized. They directly confirmed this thermal time constant during measurements to calibrate the 
detector temperature offset from the usual temperature monitoring point at the thermal control sensor/heater
assembly. This was done using a temperature sensor permanently mounted to the SAPHIRA leadless chip carrier
ceramic (this sensor
cannot be powered up during science observations because the glow levels are prohibitive). For the temperature 
offset calibration, the system was allowed to stabilize with the SAPHIRA powered down (except for the 
temperature sensor) to establish the offset between the two temperature sensors. The SAPHIRA was then powered
on and clocked normally. After it stabilized, which took approximately two hours, the on-chip sensor was used
to measure the temperature offset.

\section{Subarray Operation and Calculation of Frame Rate}\label{sec:subwindows}
There are 10 32-channel column blocks across the SAPHIRA, and it can be subarrayed to read
integer numbers of column blocks (i.e. the frame size can be adjusted by 32-pixel increments in the x
direction). On the other hand, the detector will read any number of pixels in the y direction. Multiple
subarrays can be supported at once. The 32 outputs on the SAPHIRA read out 32 adjacent pixels in a 
row at a time.
Greatly increased frame rates during subarray readouts are possible because the interleaved nature of the
outputs enables all of them to still be used.
To illustrate this, consider the case of a 128$\times$128 pixel subarray. A HAWAII-2RG has 2048 pixels per side
and 32 outputs (therefore 64 columns per output), or a HAWAII-4RG has 4096 pixels per side and 64 outputs 
(so again 64 columns per output). Either HAWAII array would utilize two outputs to read the 128$\times$128
subarray, whereas the SAPHIRA could read it using all 32 outputs. At identical pixel clocking rates, the 
SAPHIRA could read it 16 times faster.

The SAPHIRA frame rate scales roughly inversely with the number of pixels in the window. This rule is 
approximate because the array must be reset and there are extra clock cycles at the ends of rows and frames; 
the full equation for duration of one frame when operating in read-reset mode is given by 
Equation~\ref{eqn:timage-rr}.
\begin{equation}
t_{img,rr} = t_{fd} + t_{fc} + n_{r}(t_{rd}n_{cb} + t_{rst} + t_{prst} + t_{rc} + t_{ra})
\label{eqn:timage-rr}
\end{equation}
where $t_{img,rr}$ is the time to collect one image in read-reset mode and the other parameters and
their typical values are summarized in Table~\ref{table:timings}.
These clock pulses are all necessary for the detector to operate reasonably and
cannot be set to 0; the acceptable ranges are specified in the SAPHIRA manual. 
Read-reset-read mode is similar, but each pixel is read twice, and there is an 
extra $t_{rc}$ per row.
\begin{equation}
t_{img,rrr} = t_{fd} + t_{fc} + n_{r}(2t_{rd}n_{cb} + t_{rst} + t_{prst} + 2t_{rc} + t_{ra})
\label{eqn:timage-rrr}
\end{equation}
For sample up-the-ramp mode, the time to collect a frame (not including the time to reset, 
since that typically is not done every frame) is
\begin{equation}
t_{img,sutr} = t_{fd} + t_{fc} + n_{r}(t_{rd}n_{cb} + t_{rc})
\label{eqn:timage-utr}
\end{equation}
Naturally, the framerate $f$ is
\begin{equation}
f = t_{img}^{-1}
\label{eqn:framerate}
\end{equation}
The timing values for our current 1 MHz pixel rate are summarized in Table~\ref{table:timings}. These 
are subject to tuning but can be used as typical values. In line-by-line reset modes, these enable a 
1.69
kHz frame rate for a 
128$\times$128 pixel subarray or 339
Hz frame rate for 320$\times$256 pixel full frames.
\begin{table}[h!]
\centering
\begin{tabular}{ |c|c|c|c| }  \hline
\textbf{Parameter} & \textbf{Definition} & \multicolumn{2}{|c|}{\textbf{Value}} \\ \hline
 &  & \multicolumn{2}{|c|}{Frame size} \\
 \cline{3-4}
 & & 128$\times$128 px & 320$\times$256 px \\ \hline
$t_{fd}$ & duration of frame demand clock pulse & 310 ns & 310 ns \\ \hline 
$t_{fc}$ & duration of frame completion clock pulse & 410 ns & 410 ns \\ \hline 
$n_{r}$ & number of rows being read & 128 & 256 \\ \hline 
$t_{rd}$ & time that an output spends on each pixel & 1 $\mu$s & 1 $\mu$s \\ \hline 
$n_{cb}$ & number of 32-pixel column blocks in the image & 4 & 10 \\ \hline 
$t_{rst}$ & time to reset a row & 100 ns & 1 $\mu$s \\ \hline 
$t_{prst}$ & post-reset delay & 400 ns & 400 ns \\
 (global resets)  & &  &  \\ \hline
$t_{prst}$ & post-reset delay & 120 ns & 120 ns \\ 
 (line-by-line resets)  & &  &  \\ \hline
$t_{rc}$ & duration of row completion clock pulse & 210 ns & 210 ns \\ \hline
$t_{ra}$ & duration of row advance clock pulse & 210 ns & 210 ns \\ \hline
\end{tabular}
\caption{A summary of the various times for a full-frame and 128$\times$128 subarray readout of the 
SAPHIRA. These times can be used in Equations~\ref{eqn:timage-rr}-\ref{eqn:framerate} to calculate the expected frame rate. Except for the time to reset the detector, 
the timings are independent of window size. Except for the post-reset dead time, the timings 
are independent of reset mode.}
\label{table:timings}
\end{table}

It should be noted that in all readout modes described above, the detector is continuously clocking 
and reading
pixels. This is known as a rolling shutter; different parts of the frame are read at different
times. This can be problematic for observations for which the flux is rapidly changing and
synchronization is important. For example, in present pyramid wavefront sensors deployed on-sky, 
the beam is modulated in a circle around the tip of the pyramid in order to increase the linearity 
of the wavefront sensor's response. This causes each pupil image to be illuminated at a different
time. In SCExAO, the beam completes one full modulation per EMCCD image, and the EMCCD has a fast
frame transfer which functions as a global shutter, so the detector does not notice the pupils 
being illuminated at different times. On the other hand, a detector with a rolling shutter might
see gradients across the pupil images caused by the modulation, interpret it as an optical
aberration, and drive the deformable mirror to a suboptimal shape. The Keck SAPHIRA solves this 
challenge 
by synchronizing the start of each frame's readout to the modulation so that the part of the frame 
being read does not include the pupil which is changing in illumination. The time to modulate is
slightly greater than the time to read one image, but the dead time between reads is minimized
for the reason described in the next paragraph.

The integration times of an image given by Equations~\ref{eqn:timage-rr}-\ref{eqn:timage-utr} 
depend entirely on the 
durations of the various clock pulses and number of pixels being read. A user could insert
an additional delay in order to increase the detector's integration time. However, unless storage space
is limited, this is suboptimal. Instead of inserting a delay to make each exposure longer, 
the user does better to clock the detector at its maximum rate, increase the avalanche gain
to use the entire well depth with each integration (assuming read-reset or read-reset-read 
mode), and then average images together. This reduces the read noise and therefore improves
the signal to noise ratio. This contrasts with a normal (non-avalanche multiplying) detector,
wherein reading and resetting the array when very little target flux has accumulated and 
then coadding the exposures results in the read noises adding in quadrature and therefore 
a reduced signal-to-noise ratio. (In both normal and avalanche-multiplying detectors, 
non-destructively reading up the ramp at maximum readout rate until the entire well depth
has been used results in optimal signal to noise, assuming that ROIC glow is negligible.
For SAPHIRA, one can trade off the ramp length and avalanche gain; the optimal 
signal-to-noise ratio can be obtained by
selecting the avalanche gain that produces the minimum dark current and then using a
ramp length that fully utilizes the detector's dynamic range.)

\section{Radio Frequency Noise and Cryogenic Preamplifiers}\label{sec:rfi}
Radio frequency interference (RFI) noise is the dominant noise source for SAPHIRA's deployments
to Subaru Telescope with the SCExAO instrument, but it is negligible in other environments. 
In the laboratory, the measured SAPHIRA CDS read noise (9 $e^-$ RMS\cite{Atkinson2014} at unity 
avalanche gain and a Generation III Leach Controller\cite{Leach2000} at a 265 kHz clocking rate) is very 
comparable to the read noise of conventional HAWAII series arrays~\cite{Hall2016B, Fox2012}.
The read noise was similar in telescope environments such as the NASA IRTF and Kitt Peak. However when 
mounted to the SCExAO instrument at the Subaru telescope, the noise is increased by a factor of 4.5, and 
we have been unable to reduce this through the normal procedures. The SAPHIRA is ground isolated within 
the cryostat with the analog and digital grounds brought out separately to the controller and then to a 
``star'' ground point. The internal radiation shield and outer vacuum vessel are tied separately to this
same star ground to provide concentric Faraday cages. The entire camera system is isolated from the 
SCExAO bench except through a tie from the star ground to the primary SCExAO ground. 
SCExAO is characterized by an extremely challenging RFI environment and nearly all the instruments experience
higher levels of noise than they did during laboratory testing elsewhere.

\begin{figure}[!ht]
\begin{centering}
\includegraphics[width=15 cm]{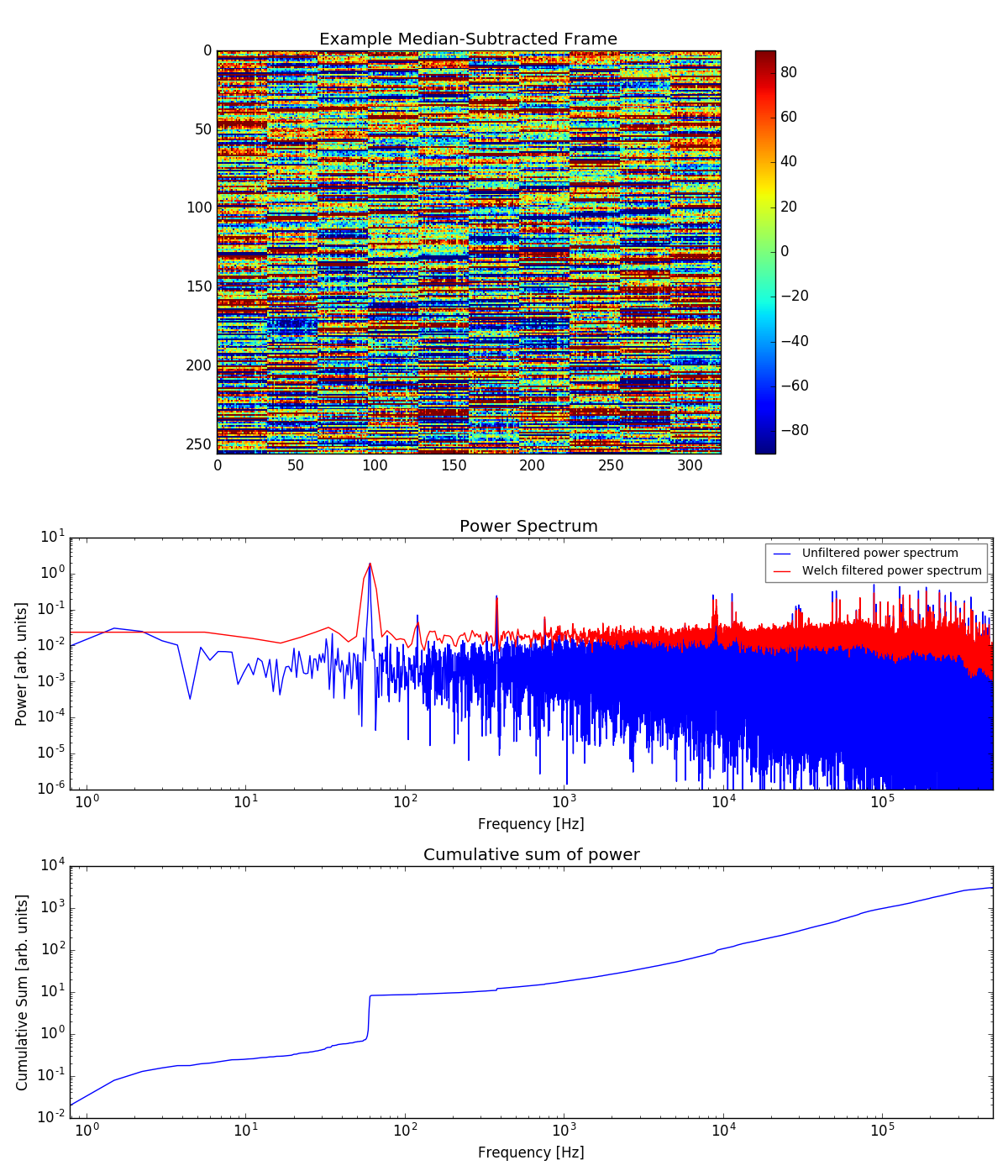}
\caption{Shown at top is a sample single frame (with the pedestal noise subtracted) which exhibits
the radio frequency interference of the SCExAO environment. In the middle is a power spectrum 
calculated from the corresponding full image 
cube. Apart from a major spike at 60 Hz (likely corresponding to ground contamination from
other electronics), the noise is largely white. Plotted over the power spectrum is a 
Welch-filtered version. The Welch filter~\cite{welch67} reduces noise in a power spectrum, but this 
comes at the
expense of spectral resolution. As a result, the lowest frequency bins of the Welch-filtered
data are much broader. At the bottom is the cumulative sum of the unfiltered power spectrum.}
\label{fig:rfi}
\end{centering}
\end{figure}

RFI shows up as 32-pixel-wide blocks that oscillate in value (Figure~\ref{fig:rfi}). A 
power spectrum can reveal some of the sources of RFI.
In our current SCExAO deployment, the noise is 
white except for a 60 Hz spike. In previous deployments, it was dominated by a few discrete 
frequencies (and therefore could be mostly filtered out from the data). The RFI changes with time, 
grounding setups, and the arrangement of wires on the outside
of the camera. We found that RFI was decreased by increasing the shielding of cables, using twisted-pair
cables, eliminating ground loops, and using a ``clean'' (i.e. with minimal other electronics connected)
ground point. RFI behaves as read noise in the noise comparisons of Section~\ref{sec:comparison}.

We are presently testing cryogenic preampliers that were developed at Australia National University. These
sit near the detector inside the camera and amplify its electrical signals prior to the wires to the readout
electronics. We anticipate that these will greatly reduce the RFI noise. Additionally, our maximum pixel rate
until now has been limited to about 1 MHz per output due the ROIC output drivers causing settling issues with
our $\sim$1 $m$ cable lengths. We expect that the preampliers will enable faster readout rates.

\section{Avalanche Gain, Charge Gain, and the Excess Noise Factor}~\label{sec:gains}
%
%
In a conventional source-follower HgCdTe array such as the Teledyne HxRG family~\cite{Beletic2008} 
or the Raytheon Virgo~\cite{Starr2016}, there are two gains are of interest: the voltage gain (which has 
units of $\mu V$/ADU) and the charge gain (which has units of $e^-$/ADU). The charge
gain is related to the voltage gain through $Q = C V$ where $C$ is the integrating node capacitance 
(the sum of the photodiode and ROIC capacitances) and has a typical value of 20 to 40 $fF$. The
voltage gain is a property only of the ROIC. In the previously mentioned infrared detector arrays, the
bias voltage is low enough that there is minimal change in the photodiode capacitance as the pixel 
accumulates photons, and this effect can be compensated for with a linearity correction.
However, the
situation is somewhat more complicated for SAPHIRA. The user adjusts the bias voltage across the 
detector over a 20 $V$ range in order to set
the avalanche gain. At bias voltages too low to produce any avalanche gain ($\lesssim$1 $V$), operation
is similar to the conventional source-follower arrays. However, as the bias voltage
of SAPHIRA increases, avalanche multiplications begin to occur and the diode capacitance, and therefore
the charge gain, decreases.
Both the avalanche multiplication and the decreasing charge gain should be taken into account when 
converting between collected photons, electrons, and ADUs for data from SAPHIRA detectors.

During SAPHIRA operation, the input node of each pixel is reset to the selected bias voltage relative to
the detector common
voltage. After the reset is lifted, photoelectrons discharge the integrating node, and the pixel 
saturates when the node voltage equals common. The node voltage is read out by a ROIC with 
a typical gain of 0.8 to 0.9 (for the single source follower design of the ME1000/ME1001)
and is 
further amplified by the controller preamplifier chain before conversion to ADU. The voltage gain of the 
system (in units of $\mu V/$ADU) can be calibrated by varying the reset voltage in known increments and
measuring changes in the corresponding digitized signal.
The charge gain is normally determined through the signal vs. variance 
method, although Finger et al.\cite{Finger2005} directly measured a $V$ vs $Q$ relation for an H2RG
detector by connecting it to a calibrated large external capacitor and measuring voltage changes in each
while pixels were exposed and then reset.

\begin{figure}[!ht]
\begin{centering}
\includegraphics[width=12 cm]{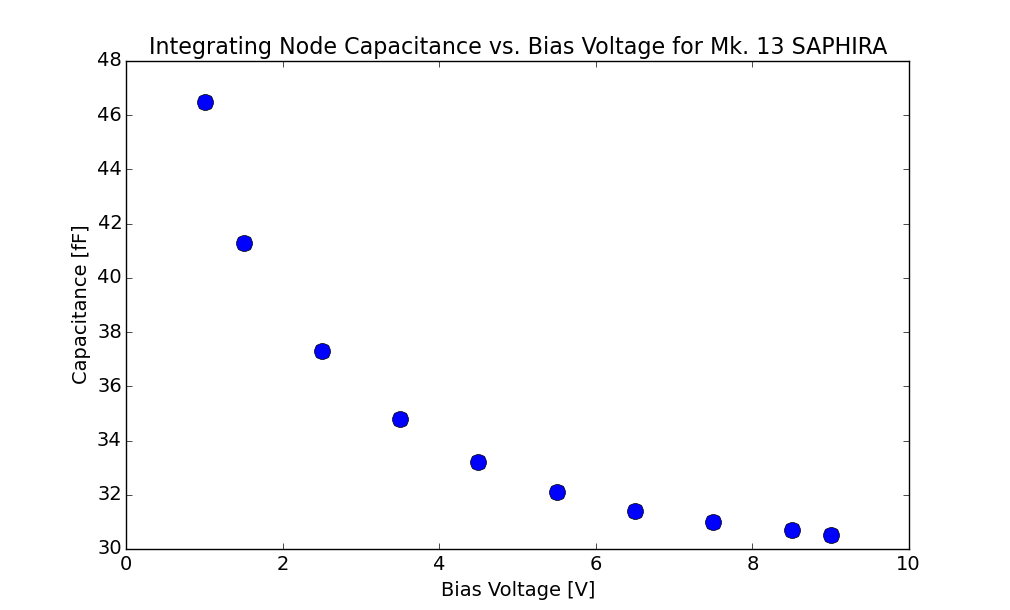}
\caption{The measured integrating node capacitance vs. bias voltage for a Mk. 13 SAPHIRA detector. The
data are from Dr. Ian Baker (private communication). 19.9 $fF$ of the integrating node capacitance is
due to the silicon readout, and the remainder is the diode. It is important to take into account the 
increasing capacitance at low bias voltages when measuring avalanche gains.}
\label{fig:capacitance}
\end{centering}
\end{figure}

The photodiode capacitance of SAPHIRA increases as the node is discharged, 
and this causes a flux vs. time plot to depart from linearity as the pixel accumulates flux.
Figure~\ref{fig:capacitance} depicts the photodiode capacitance of a Mark 13 SAPHIRA from 1 $V$ (where
the avalanche 
gain is negligible but the capacitance varies significantly with bias voltage) to 9 $V$ 
(where the diode is fully depleted so the capacitance is insensitive to bias, but the avalanche gain is 
$\sim$5). 
Avalanche gain becomes appreciable at a bias voltage of 2-3 $V$, and by biasing the detector by up to 20 
$V$, avalanche gains of $\sim$600 can be obtained. In all cases, the unit cell source
follower still limits dynamic range to a few hundred $mV$. 
The avalanche gain is usually measured by varying the bias voltage under constant detector illumination 
and defining that the flat portion of the response curve (which occurs at 1-2.5 $V$ bias) corresponds to
an avalanche gain of 1.
From Figure~\ref{fig:capacitance}, it is evident that for avalanche bias voltages $\lesssim$10 $V$, the 
change in capacitance with bias must be included in this calculation.

Read noise and RFI noise are independent of avalanche gain, so one can improve the signal to noise
ratio of observations by amplifying the signal of photons with a higher avalanche gain. However, as
avalanche gain increases, the dynamic range (in photons) of the
detector is reduced because the well depth in electrons remains constant, but more electrons are produced
per photon. At sufficiently high bias voltages, tunneling current (a type of dark 
current) increases dramatically~\cite{Atkinson2017}, can dominate over the read 
noise, and thereby can degrade the signal to noise ratio of observations~\cite{Finger2016}.

The excess noise factor $F$ characterizes noise that occurs during the multiplication process of
avalanche photodiodes.
$F$ is the ratio of the signal to noise of the photon induced charge to that of the avalanche-multiplied
charge~\cite{McIntyre1966}; noise-free amplification produces $F=1$.
The advantage of HgCdTe over other semiconductors is that only electrons participate in the avalanche
process, and this has the potential for noise-free amplification. Finger et al.~\cite{Finger2016} measured
an excess
noise factor of $F=1$ at 60 K at all gains and a maximum of $F=1.3$ at an avalanche gain of 421 at 90 K.

\section{Conclusions}
For most AO-related observations, we operate SAPHIRA in read-reset mode, wherein each row is read 
before being reset, and the pedestal noise of the detector is removed by subtracting
a reference dark frame. This mode enables the maximum usable frame rate and regular time sampling. 
When a minimum of noise is desired and long periods between detector resets are acceptable, the SAPHIRA
can be operated in up-the-ramp mode, wherein there are many reads between resets. The user then
fits a line to the flux of each pixel over time, which reduces read noise. If a pixel saturates, one
can estimate the correct flux by fitting only to the reads that occurred before the saturation. 
Read-reset-read mode, wherein a row was read, reset, and read again, and this repeats for
the next line, failed to deliver the hoped-for low noise due to a post-reset settling
effect that caused spurious flux in the columns read in the first few microseconds after
resets. For this reason, we do not commonly use read-reset-read mode. 

Because the pixel rate is determined by the readout electronics (and therefore relatively constant) 
and the outputs are designed to read contiguous 32-pixel blocks,
higher frame rates can be achieved by reading $32x \times y$ pixel subarrays, where $1 \leq x \leq 10$,
and $1 \leq y \leq 256$. For example, at a 1 MHz clocking rate in read-reset mode, one can read the 
full 320$\times$256 pixel frame at a rate of about 339 Hz or a 128$\times$128 subarray at about 
1690 Hz. The detector itself is capable of 10 MHz pixel rates, and ESO operates their SAPHIRA
arrays at typically 5 MHz rates.


SAPHIRA detectors enable the potential for new observing modes due to their high frame rate,
low noise, and infrared sensitivity. We hope that this paper can assist users in
utilizing the detectors to their maximum potential. 

\acknowledgments 
The authors acknowledge support from NSF award AST 1106391, NASA Roses APRA award NNX 13AC14G, and the 
JSPS (Grant-in-Aid for Research \#23340051 and \#26220704). Sean Goebel acknowledges funding support
from Subaru Telescope and the Japanese Astrobiology Center.
\bibliography{report} 
\bibliographystyle{spiebib} 

\end{document}